\documentstyle[psfig,aprm2002]{article}
\title{%
The UV (GALEX) and FIR (ASTRO-F) All Sky Surveys 
\/:\\ the measure of the dust extinction in the local universe}
\authors{
Buat, V.\affilmark{1},
Takeuchi, T.T.\affilmark{2},
Boselli, A.\affilmark{1},
Burgarella, D.\affilmark{1}
Hirashita, H. \affilmark{3},
Tomita, A.\affilmark{4},
Shibai, H.\affilmark{5},
Milliard, B. \affilmark{1},
Donas, J. \affilmark{1},
Yoshikawa, K. \affilmark{6},
Inoue, A.K. \affilmark{7}
and
Tajiri, Y.Y. \affilmark{7}
}

\affiltexts{
\affiltext{1}{Laboratoire d'astrophysique de Marseille, France},
\affiltext{2}{National Astronomical Observatory, Tokyo, Japan},
\affiltext{3}{Osservatorio Astrofisico di arcetri, Firenze, Italy},
\affiltext{4}{Faculty of Education, Wakayama, Japan},
\affiltext{5}{Nagoya University, Japan}
\affiltext{6}{University of Tokyo, Japan}
\affiltext{7}{University of Kyoto, Japan}

}

\firstauthor
{VB}
{veronique.buat@oamp.fr}

\papernoADS{
PL1-1
}
\authorsADS{%
Buat, V.;
Takeuchi, T.;
Boselli, A.;
Burgarella, D.;
Hirashita, H.;
Tomita, A.;
Shibai, H.;
Milliard, B.;
Donas, J.;
Yoshikawa, K.;
Inoue, A.;
Tajiri, Y.;

}
\affiliationsADS{%
AA(LAM, France)
AB(NAO Japan)
AC(LAM, France)
AD(LAM, France)
AE(Arcetri, Italy)
AF(Faculty of Education, Japan)
AG(Nagoya University)
AH(LAM, France)
AI(LAM, France)
AJ(University of Tokyo)
AK(Kyoto University)
AL(Kyoto University)
}

\begin{document}

%
\begin{abstract}
Before the end of 2002 will be launched the GALEX satellite (a NASA/SMEX 
project) which 
will observe all the sky in Ultraviolet (UV) through filters at 1500 and 2300 
$\AA$ 
down to m(AB)$\sim 21$.\\
In 2004 will be launched the ASTRO-F satellite which will perform an all sky 
survey at Far-Infrared (FIR) wavelengths.\\
The cross-correlation of both suveys will lead to very large samples of galaxies 
for which FIR and UV fluxes will be available. Using the FIR to UV flux ratio as 
a quantitative tracer of the dust extinction we will be able to measure the 
extinction in the nearby universe (z$<$0.2) and to perform a statistically 
significant analysis of the extinction as a function of galactic properties.\\
Of particular interest is the construction of pure FIR and UV selected samples 
for which the extinction will be measured as templates for the observation of 
high redshift galaxies. 
\end{abstract}

%
\section{Introduction}

The problem of the dust extinction in galaxies is crucial for the study of 
galaxy formation and evolution. Indeed the light emitted by the stars is 
partially absorbed by the dust before escaping galaxies: the determination of 
the true stellar content and of the star formation activity inside galaxies can 
only be done once the amount of extinction is known. The situation is dramatic 
at ultraviolet wavelengths which are observed in the visible at high redshift 
and where the extinction is particularly severe (e.g. Meurer et al. 1999, 
Steidel et al. 1999).

\section{How to estimate the UV dust extinction}

Our knowledge of the UV extinction by dust internal to galaxies comes from 
the studies of very limited samples of galaxies in the nearby universe.  Two 
major 
methods have been proposed based on the analysis of
\begin{enumerate}
\item the shape of the UV continuum between 1200 and 2500 $\AA$ (Calzetti et al. 
1994)
\item  the $\rm F_{FIR}/F_{UV}$ ratio and an energetic budget (Buat \& Xu 1996, 
Meurer et al. 1999)
\end{enumerate}
\subsection{Using the shape of the UV continuum}

This method has been built for central starbursts in galaxies observed by IUE 
(Kinney et al. 1993). It is based on the fact that the UV energy distributions 
for young starbursts are very similar. They can be fitted by a power-law of the 
form $\rm F_{\lambda} = \lambda^{\beta}$ (Leitherer \& Heckman 1995) and changes 
in the exponent $\beta$  is attributed to reddening (Calzetti et al. 1994, 
Meurer et al. 1995). The method was originally calibrated with the measure of 
the Balmer decrement (Calzetti et al. 1994) and new calibrations account for the 
far-infrared emission of the galaxies (Calzetti et al. 2000). Several recipes to 
estimate the extinction in the UV range have been deduced from this approach 
(Meurer et al. 1999, Calzetti et al. 2000).
 Recent studies have  criticized  the universality
 of the link between  the slope $\beta$ and the dust extinction: theoretical 
  studies explain the close link between the extinction and the slope $\beta$ 
for starburst galaxies  only under restrictive conditions (Witt \& Gordon  
2000);
   moreover the 
 $\beta$-$\rm F_{FIR}/F_{UV}$ relation observed for starburst galaxies does not 
hold for Ultra Luminous Infra Red Galaxies (ULIRGs, Goldader et al. 2002)
 or normal spirals (Bell 2002). 
 
\subsection{Performing an energetic budget}

The principle of this approach is very simple: the stellar emission is absorbed 
by the dust which re-emits the energy in the far-infrared. Since the UV non 
ionizing flux ($\rm 912 \AA <\lambda<\sim 3000 \AA$ ) is the main source of dust 
heating in galaxies forming stars actively  the $\rm F_{FIR}/F_{UV}$ ratio is a 
powerful tracer of the dust extinction (Buat \& Xu 1996, Meurer et al. 1999, 
Witt \& Gordon 2000). Indeed the  $\rm F_{FIR}/F_{UV}$ ratio can be calibrated 
quantitatively in terms of extinction: it is relatively insensitive to the geometry 
of the dust versus that of the stars, to the dust properties and to the details 
of the star formation history (Gordon et al. 2000). \\
In order to apply this method one must estimate the total dust emission from the 
mid-infrared to the sub-mm. A good estimate of this total dust emission for 
normal galaxies needs 
observations at long wavelengths ($\lambda > 60 \mu m$, Dale et al. 2001).

\section{The UV and FIR observations}

\subsection{The existing data}

Until now the UV data are relatively scarce: wide field imagers  ($\lambda \sim 
150-200 $ nms) 
have only covered  limited sky areas (Donas et al. 1987 (SCAP), Milliard et al. 
1994(FOCA), Deharveng et al. 1994 (FAUST),  Stecher et al. 1997(UIT)) and the 
infrared 
data for statistical studies came essentially from IRAS. The low 
sensitivity of IRAS has limited the sample of galaxies with photometric data in 
FIR and UV  to few hundreds  objects: only 
5$\%$  of the galaxies detected in the UV survey of the FOCA experiment 
of 70 deg$^2$ have a FIR counterpart and the resulting sample contains only 80 
galaxies (Buat et al. 1999).\\
 The situation is even worse for UV spectroscopy since we still rely on the 
observations of the IUE satellite (Kinney et al. 1993) complemented with some 
data from HUT (Leitherer et al. 2002). The comparison with IRAS data is very 
difficult given the small aperture of the IUE telescope (10 arcsec $\times$ 20  
arcsec) as compared to the optical size of the galaxies often as large as 
several  arcmins (Meurer et al. 1999, Buat et al. 2002).\\
If visible spectroscopy is needed in complement to UV and FIR observations the 
samples are reduced to few 
tens of galaxies (Meurer et al. 1999, Buat et al. 2002).\\

 \subsection{Future observations}
The situation will be greatly improved with the launch of all sky surveyors in 
UV (GALEX) and FIR (ASTRO-F): we expect several millions of galaxies with both 
FIR and 
UV data. \\
The launch of the UV satellite GALEX is scheduled for the end of 2002: it  will 
perform a full sky imaging survey at a limiting flux (5$\sigma$) of 15.5 
$\mu$Jy at 1500 \AA\ and 8.5 $\mu$Jy at 2300 \AA. 
A deeper survey covering 160 deg$^2$ will reach 0.1 $\mu$Jy in both bands.
Moreover, the GALEX survey will provide low resolution (R=100) UV spectra of 
thousands of galaxies from 1350 to 3000 $\AA$. \\
The ASTRO-F All Sky Survey sheduled for 2004 is expected to be ten times deeper 
than the IRAS one.
The  limiting flux at 80 $\mu$m (FIS instrument, wide-S filter) will reach 16 
mJy  
(http://www.ir.isas.ac.jp/ASTRO-F).\\
The performances of both surveys can be compared in figure 1.
In this figure we have plotted  the FIR flux versus the UV one.  
Both fluxes are defined as $\nu\cdot f_{\nu}$. The expected limits for GALEX at 
1500 \AA (FUV) and 2300 $\AA$ (NUV) are reported  as vertical dashed lines 
and right arrows. 
The  limiting flux at 80 $\mu$m of ASTRO-F is 
 represented as an horizontal dashed line and arrows up. For comparison we have 
also reported the limit of IRAS {\sl Faint Source Catalog} at 60 microns. Observed data for existing 
samples of galaxies are also plotted (see the legend of the figure for details).
Diagonal dotted lines represent the locus of constant $\rm F_{FIR}/F_{UV}$.
Most of the galaxies in the nearby universe have a $\rm F_{FIR}/F_{UV}$ ratio   
comprised between 0.1 and 100 
 (Buat \& Xu 1996, Buat et al. 1999). 
\begin{figure*}[ht]
\begin{center}
\leavevmode\psfig{angle=-90,file=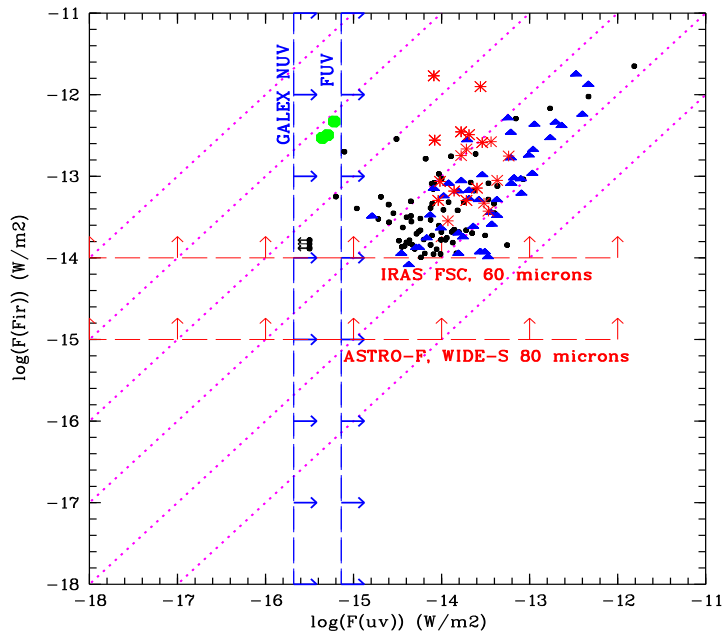,height=90mm,width=140mm}
\end{center}
\caption{The expected limits of the all sky surveys of ASTRO-F and GALEX. The 
IRAS-FSC/FOCA  sample (Buat et al. 1999) is 
  plotted with black dots; the IUE sample and the sample of nearby galaxies with 
visible spectroscopy are plotted with red stars and blue triangles respectively 
(Buat 
et al. 2002); 3 ULIRGs (Trentham et al. 1999) are plotted with green filled 
circles. The 
diagonal dotted lines are the locus of constant FIR to UV 
  flux ratio ($\log(F_{\rm FIR}/F_{\rm UV})=-1,0,1,2$). 
  The detection limit of ASTRO-F and IRAS/FSC are  indicated by  horizontal 
  dashed lines and  arrows up, the detection limits of GALEX  are indicated by 
vertical dashed lines and right 
  arrows.}
\label{fig:1}
\end{figure*}
\subsection{Statistical analyses with the future GALEX and ASTRO-F data}

A first look to the figure 1 leads to the conclusion that the two surveys are 
very well suited: the  full cross-correlated GALEX--ASTRO-F sample is expected  
to be not strongly 
biased against or towards galaxies with a low or high extinction.
Therefore this full sample can be used for statistical studies of the FIR and UV 
properties of the galaxies in the local universe in terms of 
dust properties, extinction or star formation. Here are some examples of the 
topics which might be adressed with these data.
\begin{enumerate}
\item The process of dust heating by the UV photons will be studied. The dust 
temperature will be 
estimated with the ASTRO-F observations at 80 and 155 $\mu$m. 
\item For each type of galaxies, we will compare both indicators of dust 
extinction: the slope $\beta$ of the UV 
continuum (with the two bands of GALEX at 1600 and 2300 $\AA$) and the $\rm 
F_{FIR}/F_{UV}$ ratio. Complementary data in the visible will be available from 
the large surveys SDSS or 6dF.
\item 
The high sensitivity of  ASTRO-F will  provide
us with a wide range of luminosity of galaxies in the local universe.
Hence, we will obtain a large sample of dwarf galaxies, whose FIR 
properties have been still only poorly understood.
Dwarfs generally have low metallicities, so it also means that we will be
able to study the dust extinction of low-metallicity objects unobserved in FIR 
until now.
\item
The biases induced by the adoption of UV  or FIR  selection criteria in 
 observations of galaxies at any redshift will be properly analysed and 
understood by studying the 
statistical properties of galaxies in pure UV and FIR selected subsamples 
drawn from the GALEX--ASTRO-F sample. For example from the  Figure 1 we can 
build a  UV selected sample for which a FIR detection with ASTRO-F is very 
likely (and for which a non detection will put a strong constraint)  by 
truncating the 
GALEX All Sky Survey  at $\log 
(f_{\rm uv}) \sim -14.5 $  (equivalent to m(AB) = $\sim 18$). 
We expect $\sim 400 000$ objects for such a  sample.
\item
An estimate of the total star formation rate in the local universe will be 
obtained by summing directly the star formation rate from the observed UV and 
FIR emissions: $\rm SFR (total) = SFR (UV_{obs}) +SFR (FIR_{obs})$ (Flores et 
al. 1998, Buat et al. 1999)
\end{enumerate}

\section*{References}

\reference
Bell, E. 2002, astroph/0205439 
\reference
Buat, V., Xu, C. 1996, A\&A, 306, 61
\reference
Buat, V., Boselli, A., Gavazzi, G., Bonfanti, C. 2002, A\&A, 383, 801
\reference
Calzetti, D., Kinne,y A.L., Storchi-Bergmann, T. 1994, 
ApJ, 429, 582
\reference
Calzetti, D., Armus, L., Bohlin, R.C., Kinney, A.L.,      
Koornneef, J., Storchi-Bergmann T. 2000, ApJ, 533, 682
\reference
Dale, D., Helou, G., Contursi, A., Silberman, N.A., Kolhatkar, S. 2001, ApJ, 
 549, 215
\reference
Donas, J, Deharveng, J.M., Milliard, B., Laget, M., 
Huguenin, D. 1987,  A\&A 180, 12
\reference
Goldader, J.D., Meurer, G.R., Heckman, T.M., Seibert, M., Sanders, D.B., 
Calzetti, D., Steidel, S.C. 2002, ApJ 568, 651
\reference
Deharveng, J.M., Sasseen T.P., Buat, V., Bowyer, S., 
Wu, X. 1994, A\&A 289, 71
\reference
Flores, H., Hammer, F.,
 Thuan, T. X., Cesarsky, C.,
 Desert, F. X., Omont, A., Lilly, S. J.,
 Eales, S., Crampton, D.,
 LeFevre, O. 1998, ApJ, 517, 148
\reference
Gordon, K., Clayton, G., Witt, A., Misselt, K. 2000, 
ApJ, 
533, 236
\reference
Kinney, A., Bohlin, R., Calzetti, D., Panagia, N., Wyse, 
R. 1993, ApJSS, 86, 5
\reference
Leitherer, C.,  Heckman, T.M. 1995, ApJS 96, 9
\reference
Leitherer, C., Li, I-H, Calzetti, D., Heckman, T.M.  
2002, ApJSS, submitted
\reference
Meurer, G.R., Heckman, T.M., Leitherer, C., Kinney 
A.L., Robert, C., Garnett, D.R. 1995, AJ, 110, 2665 
\reference
Meurer, G.R., Heckman, T.M., Calzetti, D. 1999, 
ApJ, 521, 64
\reference
Milliard, B., Donas, J., Laget, M., Huguenin, D. 
1994, {\it The balloon-borne 40-cm UV imaging telescope FOCA; results and 
perspective}, 11th ESA Symposium, Montreux
\reference
Stecher, T.P., Cornett, R.H., Greason, M.R., Landsman, 
W.B., Hill, J.K., Hill, R.S. et al. 1997, PASP, 109, 584
\reference
Steidel, S., Adelberger, K., Giavalisco, M., Dickinson, M., Pettini, M.  1999, 
ApJ, 519, 1
\reference
Witt, A. N., Gordon, K. D. 2000, ApJ, 528, 799
\vspace{-1mm} 

\end{document}